\documentclass{camera}
\usepackage{graphicx}  

\begin{document}

%
\title{The IceTop experiment in 2010}

%
\author{Todor Stanev}

%
\organization{Bartol Research Institute, Department of Physics and Astronomy,
 University of Delaware, Newark, DE 19716, U.S.A. for the IceCube collaboration}

\maketitle

\begin{abstract}
  We present the current status of the IceTop air shower array
 on top of the IceCube neutrino detector that IceTop can use 
 as a huge detector of TeV muons. We laos give a brief discussion
 of different types of air shower events that contain information
 on the spectrum and composition of the cosmic rays in a wide 
 energy range.    
\end{abstract}

%

\section{Introduction}

 IceTop is the air shower array on top of the IceCube neutrino
 telescope at South Pole. It was proposed to support IceCube
 in terms of relative
 and absolute pointing, to guard against high energy air showers
 being misreconstructed as high energy neutrinos and to perform 
 cosmic ray physics research. This includes:\\
 \ \ \ \ $\bullet$ measurement of the cosmic ray spectrum using
 the surface detectors of IceTop\\
 \ \ \ \ $\bullet$ measurements of the cosmic ray spectrum and
 composition by studies of the angular distribution of the
 muon bundles InIce\\
 \ \ \ \ $\bullet$ measurements of the spectrum and composition
 of cosmic rays in coincident IceTop/IceCube events

  The IceCube tedector is deployed at a depth of 1,500 meters in ice.
 The surface energy of a verticle muon that reaches the top of IceCube
 should be 460 GeV. The surface threshold energy for a muon
 to reach the top of InIce with energy 100 GeV, and to generate a
 signal in it, 
 is 626 GeV. IceCube can not measure the number of muons that reach
 it. However, it measures the energy released by these muons in their
 propagaqtion in 1 km of deep clear ice, which carries valuable information
 about the shower energy stored in high energy muons.

  The surface detectors of IceTop are plastic Cherenkov tanks of radius 1 m
 with diffusive reflective inside liner. The depth of ice in the
 tanks is 90 cm. They are covered with a black layer and tank cover.

 Eack tank is equipped with two digital optical modules (DOMs)
 that look down in the tank and are identical to those deployed
 in the deep ice. One of the photomultipliers of the
 tank is run in high gain and the other in low gain to increase
 the dynamic range of the tank. Note that the shower gamma rays
 often convert in the tanks which measure the energy flow of
 air shower electromagnetic component plus the contribution of
 GeV muons.  

   An IceTop station consists of two tanks at a distance about
 10 meters from each other and the corresponding IceCube string.
 Coincidental hits in both tanks (hard local coincidences) are 
 considered air shower events and participate in the triggering
 of IceTop which requires 6 triggered DOMs. Single tank hits are
 considered single particles: muons electrons or gamma-rays.
 Such events are called soft local coincidences and are also
 collected by the data acquisition system.
 
  The DOMs contain 10'' Hamamatsu R7081 PMT with onboard HV module
 and onboard digitization performed by 2 300 MHz ATWD, each with
 three channels with different gains. The calibration of the DOMs 
 is done by the measurement of muon signals which goes continuously
 throuout the year and affects the difference of the signals in
 the high gain and low gain DOMs as well as the sensitivity of
 different tanks and stations.

 \section{The IceTop air shower array}

\begin{figure}[thb]
\centerline{\includegraphics[width=8truecm]{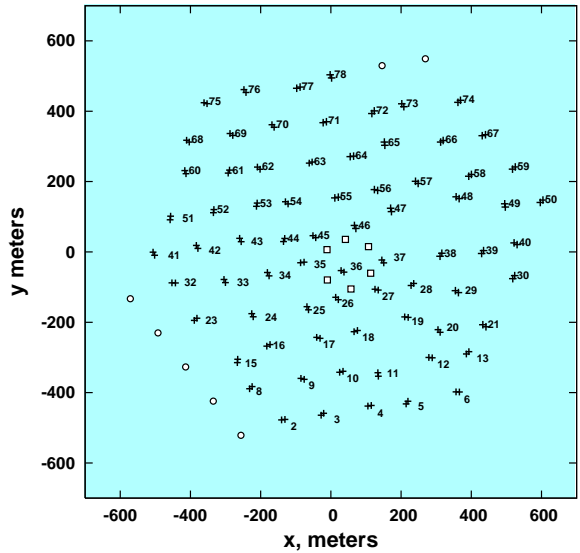}}
\caption{Map of the IceTop stations deployed by the end of the 2009/10
 season. Each tank is separately shown together with the station number.
 Seven stations are not yet deployed and are shown with circles.
 The Deep Core strongs are marked with empty squares.}
\label{it73} 
\end{figure}

  The proposed IceTop array was supposed to consist of 80 stations
 in a triangular arrangement at an average distance of 125 m from
 each other~\cite{TKG_Merida}. The total area within the perimeter of
 the array will be 1 km$^2$. The currently deployed 73 stations are
 shown in Fig.~\ref{it73}. The rest of the stations will be deployed
 in the 2010/11 South Pole season. The five stations at negative $x$
 and $y$ will be deployed as shown on the graph. The two missing stations
 at positive $x$ and $y$, which are shown on top of previous 
 South Pole experiments, will most likely be deployed in the center
 of the array, close to the Deep Core strings.

  There are four types of triggers that the DAQ collects~\cite{NIM}
 and a
 fraction of which are transmitted North for analysis: These are:\\
 \ \ \ \ $\bullet$ SMT triggers: 3 and more station events\\
 \ \ \ \ $\bullet$ LMT Triggers: 8 and more station events\\
 \ \ \ \ $\bullet$ SMT triggers with InIce activity\\
 \ \ \ \ $\bullet$ InIce triggers with IceTop activity

 When fully deployed IceTop will be able to measure the cosmic ray
 spectrum and composition between 10$^{14}$ and 10$^{18}$ eV.
 At the higher energy end its results will coincide with the lower
 energy extensions of the Southern Auger Observatory and the Telescope
 Array. One of the advantages of IceTop is that the South Pole is 
 at high altitude and the average atmoepheric depth is less than
 700 g/cm$^2$.

\section{Physics, problems, and results}

  IceTop has completed~\cite{Klepser,Kislat} a determination of
 the cosmic ray energy spectrum from the data of IceTop26, a stage
 of the experimenty with only 26 stations deployed and the
 area of the air shower array was 0.094 km$^2$. Using this data set
 the position of the cosmic knee was determined at 3.1$\pm$0.3 Pev
 (1 PeV = 10$^{15}$ eV).
 At this energy the cosmic ray energy spectrum slope changes from
 E$^{-2.71}$ to E$^{-3.11}$. 

  The current analysis concentrates on the following phases of
 the air shower array with 40, 59 and the current 73 stations
 deployed. As with any new and to large degree unusual experiment,
 there are some problems that take time to solve. One of them is
 the ever increasing size of the detector which requires separate
 Monte Carlo detector calculations depending on the array size and
 shape. Another one the increasing with time snow coverage
 of the tanks  that
 changes the waveform shapes and magnetide of the signals.
  
 IceTop is analyzing at least three main types of events.
 The first one are nearly vertical showers (zenith angle
 less than 35$^o$) that could be analyzed separately in
 IceTop and InIce. The small zenit angle is required so
 that the events are well contained on the surface and in 
 deep ice. The energy on the surface is determined from
 the data of IceTop, while the energy loss of the muon bundle
 is measured InIce. Using both data sets improves the 
 zenith angle and air shower core determination~\cite{TomF}. It also helps
 to eliminate events in which an air shower coincides in time
 with a high energy single muon. For air showers with 5 or more
 hard local coincidences the energy threshold for protons
 is 300 TeV and for iron showers is about 500 TeV.
 
 It is important to note that on average 53 muons of energy
 above 100 GeV would reach the top of IceCube from an vertical iron
 shower of energy 10 PeV. From a proton shower of the same energy 22
 detectable muons reach the top of IceCube. Another important
 parameter is the energy loss in propagation inside IceCube.
 One TeV muon loses 643 GeV in propagation on 1 km of ice.
 Since the average energy of muons in Fe initiated showers
 is lower that those in proton showers the Fe shower  muon bundle
 will have a stronger decrease of the energy loss with depth. 
 All this physics of the muon bundles energy loss is being
 implemented in the analysis of such events.

\begin{figure}[thb]
\centerline{\includegraphics[width=8truecm]{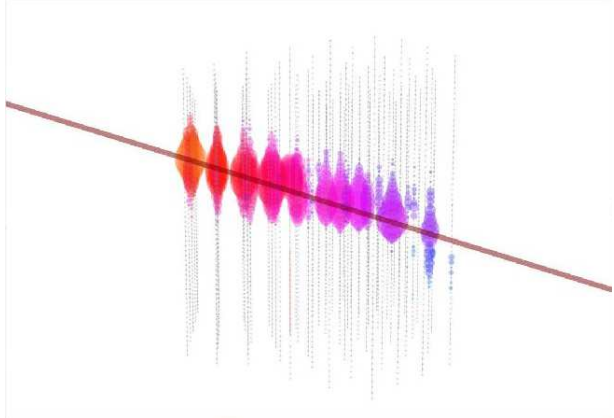}}
\caption{ The muon bundle from an inclined shower with zenith 
 angle of 72.5$^o$ detected by IceCube. This event was found by
 P.~Berghaus.}
\label{pb} 
\end{figure}

 The second type of showers is illustrated with an experimental
 event in Fig.~\ref{pb}. It shows a muon bundle of zenith angle
 72.5$^o$ detected by IceCube. This muon bundle has triggered 
 536 InIce DOMs and deposited in them almost 50,000 photoelectrons.
 The shower axis must be more than 5 km away from IceTop but it 
 still triggers a responce at the edge of the detector. 
 The energy estimate for this shower is 3$\times$10$^{18}$ eV.

 Most of the high energy events that trigger IceTop are not well
 contained within the array. When the muon bundles of these 
 showers are well contained InIce, as it is in the event shown 
 in Fig.~\ref{pb}, they can still be analyzed. The extension 
 of the muon bundle track to the surface track to the surface
 determines the position of the shower axis on the surface
 within 50 meters. Together with the information from the
 surface stations triggered in such an event the analysis is
 certainly possible. The electromagnetic and GeV muons signal
 on the surface should give a good estimate of the shower 
 primary energy and the energy released by the TeV muons InIce
 contains the information about the type of the primary cosmic
 ray nucleus. The average surface energy of a muon to be detected
 by IceCube at a zenith angle of 72.5$^o$ is 5.1 TeV and the muon
 bunch shown in Fig.~\ref{pb} contains many such muonss. 

 The third type of extensive air showers that are currently being
 analyzed are the smallest air showers that trigger only three 
 or four stations in IceTop~\cite{Bakhti}. Such events could be very
 interesting because they are in the energy range where the cosmic rays
 spectrum and composition are studied directly with balloon and
 satellite detectors. The comparison of the IceTop data to the more
 detailed experiments results will help us understand better the
 composition related air shower parameters.

 Three station events must contain only internal stations that are
 next to each other. In such case the three triggering stations 
 are in the shape of amost uquilateral triangle. Four station
 events have a diamond shape. The reconstruction of these low 
 energy showers is simpler and not as good as that of the showers
 that trigger more stations. One cannot, for example, account for
 the curvature of the shower front and has to treat is as a plain
 wave.  

 The advantage is that they correspond to a relatively small and
 narrow energy range. Monte Carlo calculations of proton showers
 with $\cos\theta$ higher than 0.9 show that the width of the 
 energy distribution that creates such showers in IceTop is
 only 0.2 in $\log_{10} E$, from 10$^{5.3}$ to 10$^{5.5}$ GeV.
 Iron showers show higher average energy and somewhat wider
 energy distribution - from 10$^{5.5}$ to 10$^{5.8}$
 GeV~\cite{Bakhti}. With increasing zenith angle the energy
 of the cosmic rays that create such showers also increases 
 and one can study the cosmic rays between 200 and 1,000 TeV
 by looking at the three and four station triggers.

 As a summery, IceTop is almost fully deployed and in less than a year
 it will have an area of 1 km$^2$. The experiment is working very well
 and the events collected during the deployment of the array are
 being analyzed. The existence of a huge muon detector as IceCube is
 makes it an air shower array that can be very successful
 in studies of the cosmic ray composition between 10$^{14}$ and 
 10$^{18}$ eV.\\[3truemm]  
{\bf DISCUSSION}\\[3truemm]
{\bf WOLFGANG KUNDT} You surprize me with the long cooling time of 
 your tanks at -25$^o$C, of order months. The rain water collectors
 of my house near Bonn take days to freeze if the temperatures fall
 below 0$^o$C. What is the osmotic contents of your water?\\
{\bf TODOR STANEV} The water quality in IceTop is very good. It is 
 filtered and well tested. There are two main reasons for the slow
 freezing of the tanks://
 1) When ice forms on the bottom and the walls of the tanks the
 freezing slows down because ice is very good insulator.\\
 2) There is a pump at the bottom of the tank that circulates the
 water in a system that deletes the air bubbles from the water.
 This pump creates heat that also slows down the freezing.



\begin{thebibliography}{99}
\bibitem{TKG_Merida} T.K.~Gaisser et al., Proc. 30th ICRC, Merida,
 Mexico (2007)
\vspace{-3truemm}
\bibitem{NIM} R.~Abbasi et al. (IceCube Collaboration), 
 NIM A{\bf 601}, 294 (2009)
\vspace{-3truemm}
\bibitem{Klepser} S.~Klepser, PhD Thesis, Humboldt-Universit\"{a}t 
 zu Berlin (2008)
\vspace{-3truemm}
\bibitem{Kislat} F.~Kislat et al., Proc 31st ICRC, Lodz, Poland (2009)
\vspace{-3truemm}
\bibitem{TomF} T.~Feusels et al., Proc. 31st ICRC, Lodz, Poland (2009)
\vspace{-3truemm}
\bibitem{Bakhti} B.~Ruzybayev et al., Proc. 31st ICRC, Lodz, Poland (2009)
\end{thebibliography}
\end{document}